\documentclass[%
 reprint,
superscriptaddress,
 amsmath,amssymb,
 aps,
pra,
floatfix,
]{revtex4-2}

\usepackage{graphicx}
\usepackage{dcolumn}
\usepackage{bm}
\usepackage{siunitx}
\usepackage[dvipsnames]{xcolor}
\usepackage{lipsum}
\usepackage{braket}
\usepackage{nicefrac}

\DeclareSIUnit\barn{barn}
\DeclareSIUnit\muN{$\mu_N$}

\begin{document}

\preprint{APS/123-QED}

\title{Hyperfine spectroscopy and laser cooling of the fermionic isotopes $^{47}$Ti and $^{49}$Ti}

\author{Jackson Schrott}
\affiliation{Department of Physics, University of California, Berkeley, CA 94720}
\affiliation{Challenge Institute for Quantum Computation, University of California, Berkeley, CA 94720}

\author{Scott Eustice}
\altaffiliation[Present address: ]{Joint Quantum Institute, National Institutes for Standards and Technology, College Park, MD 20742}
\affiliation{Department of Physics, University of California, Berkeley, CA 94720}
\affiliation{Challenge Institute for Quantum Computation, University of California, Berkeley, CA 94720}

\author{Pouya Sadeghpour}
\altaffiliation[Present address: ]{Department of Physics, University of Chicago, Chicago, IL 60637}
\affiliation{Department of Physics, University of California, Berkeley, CA 94720}
\affiliation{Challenge Institute for Quantum Computation, University of California, Berkeley, CA 94720}

\author{Rowan Duim}
\affiliation{Department of Physics, University of California, Berkeley, CA 94720}
\affiliation{Challenge Institute for Quantum Computation, University of California, Berkeley, CA 94720}

\author{Hiromitsu Sawaoka}
\affiliation{Department of Physics, University of California, Berkeley, CA 94720}
\affiliation{Challenge Institute for Quantum Computation, University of California, Berkeley, CA 94720}

\author{Dmytro Filin}
\affiliation{Department of Physics and Astronomy, University of Delaware, Newark, DE 19716}

\author{Marianna S. Safronova}
\affiliation{Department of Physics and Astronomy, University of Delaware, Newark, DE 19716}

\author{Dan M.\ Stamper-Kurn}
\affiliation{Department of Physics, University of California, Berkeley, CA 94720}
\affiliation{Challenge Institute for Quantum Computation, University of California, Berkeley, CA  94720}
\affiliation{Materials Sciences Division, Lawrence Berkeley National Laboratory, Berkeley, CA  94720}




\date{\today}

\begin{abstract}
We report on magneto-optical trapping of the two fermionic isotopes of atomic titanium, $^{47}$Ti and $^{49}$Ti.  Unlike the even mass-number isotopes, which were recently laser cooled, $^{47}$Ti and $^{49}$Ti have nonzero nuclear spins and, consequently, their atomic levels are split by hyperfine structure. Combining and comparing theoretical calculations and atomic beam-spectroscopy measurements, we determine the hyperfine structures and isotope shifts of the $\mathrm{3d^24s^2}$ $\mathrm{a^3F_4\rightarrow 3d^2(^3P)4s4p(^3P^o)}$ $\mathrm{y^5D_4^o}$ optical-pumping transition at optical wavelength \SI{391}{\nm} and the $\mathrm{3d^3(^4F)4s}$ $\mathrm{a^5F_5\rightarrow 3d^3(^4F)4p}$ $\mathrm{y^5G_6^o}$ laser-cooling transition at wavelength \SI{498}{\nm}.  With this information, we produce magneto-optical traps of both $^{47}$Ti and $^{49}$Ti by applying two additional tones of light to repump atoms to the maximum-spin states on the laser-cooling transition. Directly loading from the atomic flux of a titanium sublimation pump, we produce $^{47}$Ti and $^{49}$Ti traps with 731(190) and 1142(240) atoms, and with lifetimes of \qty{330(15)}{ms} and \qty{310(8)}{ms}, respectively.
\end{abstract}

\maketitle

Quantum simulation experiments with ultracold Fermi gases have provided numerous insights into superfluidity~\cite{regal_observation_resonance_condensation_fermionic_atom_pairs_2004}, quantum thermodynamics~\cite{nascimbene_exploring_thermodynamics_universal_fermi_gas_2010}, Fermi-Hubbard physics~\cite{greiner_quantum_2002, jordens_mott_insulator_fermionic_atoms_optical_lattice_2008} and other topics. 
Most such experiments have used the stable alkali isotopes $^{6}$Li and $^{40}$K.
Yet, limitations of these atoms have been noted in several important contexts, such as in the study of spin-orbit coupled Fermi gases~\cite{zhai_degenerate_2015}.  While ultracold Fermi gases of a few other elements have been produced, including $^{53}$Cr,  $^{87}$Sr, $^{161,163}$Dy, $^{167}$Er, and  $^{171,173}$Yb~\cite{naylor_chromium_dipolar_fermi_sea_2015, taie_su6_mott_insulator_atomic_fermi_gas_realized_largespin_pomeranchuk_cooling_2012, lu_quantum_degenerate_dipolar_fermi_gas_2012, aikawa_observation_fermi_surface_deformation_dipolar_quantum_gas_2014, chomaz_longlived_transient_supersolid_behaviors_dipolar_quantum_gases_2019,hofer_observation_orbital_interactioninduced_feshbach_resonance_yb_173_2015}, providing experimental alternatives to the alkali metals and opening up several new scientific directions, the selection of fermionic isotopes produced at ultralow temperatures remains limited.  Laser cooling new and different fermionic atoms will open up even more areas of study.

 \begin{figure*} [th]
    \centering
    \includegraphics[]{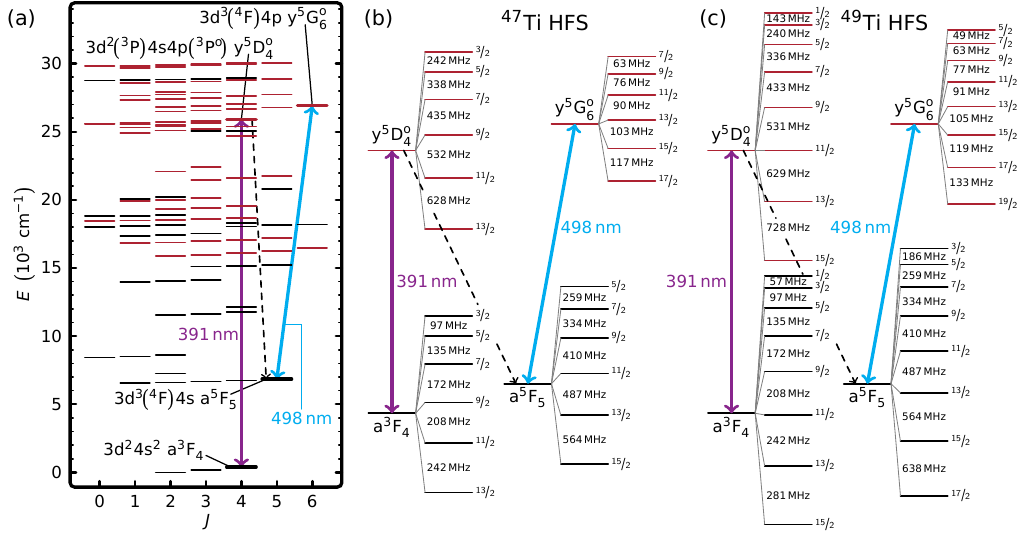}
    \caption{(a) Atomic structure of Ti, including all levels below \qty{31000}{\per\cm}. Even (odd) parity levels are shown in black (red), and the levels relevant to laser cooling are bolded and labeled. The \qty{391}{\nm} optical pumping and \qty{498}{\nm} laser cooling transitions are shown with purple and blue arrows. A dashed arrow indicates branching from the $\mathrm{y^5D_4^o}$ level to the $\mathrm{a^5F_5}$ level. (b,c) show the hyperfine structure of the four levels studied in this work for $\mathrm{^{47}Ti}$ and $\mathrm{^{49}Ti}$. The $\mathrm{a^3F_4}$ and $\mathrm{a^5F_5}$ hyperfine splittings are derived from the hyperfine constants given in~\cite{channappa_hyperfine_structure_measurements_lowlying_multiplets_47ti_49ti_59co_105pd_1965,aydin_sternheimer_free_determination_the47ti_nuclear_quadrupole_moment_hyperfine_structure_measurements_1990}, while the $\mathrm{y^5D_4^o}$ and $\mathrm{y^5G_6^o}$ splittings were determined in this work.}
    \label{fig:atom-struc}
\end{figure*}

A scheme for laser cooling several transition-metal elements, including several stable fermionic isotopes, was recently proposed~\cite{eustice_laser_2020}.  For most of these atoms, laser cooling is a two-part process in which atoms are first optically pumped into a long-lived metastable-state and then laser cooled on a cycling optical transition out of that state. The realization of this scheme for the zero-nuclear-spin and bosonic isotopes of titanium, $^{46}$Ti, $^{48}$Ti and $^{50}$Ti, was reported in Ref.~\cite{eustice_magneto-optical_2025}.  Extending this scheme to the fermionic isotopes of Ti, which have non-zero nuclear spin, requires knowledge of the hyperfine structure (HFS) of the atomic levels involved in optical pumping and laser cooling.

Here, we describe the successful laser cooling and magneto-optical trapping of the stable fermionic isotopes of atomic titanium, $^{47}$Ti and $^{49}$Ti. We present two main results.  First, we use advanced atomic structure calculations to predict (Sec.~\ref{sec:theory}), and fluorescence spectroscopy of an atomic beam to measure (Sec.~\ref{sec:spectroscopy}), the hyperfine structure and isotope shifts on the $\mathrm{3d^24s^2}$ $\mathrm{a^3F_4\rightarrow 3d^2(^3P)4s4p(^3P^o)}$ $\mathrm{y^5D_4^o}$ transition used for optical pumping, and the $\mathrm{3d^3(^4F)4s}$ $\mathrm{a^5F_5\rightarrow 3d^3(^4F)4p}$ $\mathrm{y^5G_6^o}$ transition used for laser cooling.  The hyperfine coefficients determined experimentally agree well with theory.  Second, we make use of this information to realize magneto-optical trapping of each fermionic isotope (Sec.~\ref{sec:mots}).  We find that by applying three tones of laser-cooling light -- one laser-cooling tone slightly detuned below the stretched-state resonance, and two additional repumping tones -- we produce magneto-optical traps with long trapping lifetimes.

\section{Scheme for optical pumping and laser cooling of $^{47}$Ti and $^{49}$Ti}
\label{sec:scheme}

The atomic level structure relevant to laser cooling of Ti is shown in Fig.~\ref{fig:atom-struc}.  The ground electronic configuration of Ti, $\mathrm{3d^24s^2}$, is split by fine structure into three terms, $\mathrm{a ^3F_{2, 3, 4}}$. With the fine-structure splittings in the range of $\Delta E \sim h c \times \qty{100}{\cm^{-1}}$, corresponding to thermal energies in the range of $T = \Delta E/k_B \sim \qty{100}{\kelvin}$, all these levels are well populated in an atomic beam produced by Ti sublimation, at the sublimation temperature around \qty{1400}{\kelvin}~\cite{schrott_atomic_2024}. However, none of these levels support the near-cycling optical transitions required for laser cooling.

\begin{figure}
    \centering
    \includegraphics[]{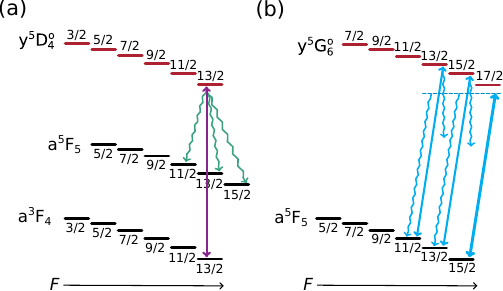}
    \caption{(a) Optical pumping on the $F=13/2\rightarrow F'=13/2$ transition in $^{47}$Ti produces population in the $F=11/2,13/2,15/2$ levels of the $\mathrm{a^5F_5}$ state. Because of the large degeneracy of the $F=13/2$ level of the $\mathrm{a^3F_4}$ state, approximately $1/4$ of the atoms in the $\mathrm{a^3F_4}$ state are addressed by this light.  (b) Laser cooling light is detuned to the red of the $F=15/2\rightarrow F'=17/2$ transition. Off-resonant Raman scattering on the $F\rightarrow F$ and $F\rightarrow F-1$ transitions out of the $F=15/2$ state leads to depumping of population into the $F=13/2,\ 11/2$ states (long curvy arrows). Repump tones resonantly drive the $F=13/2\rightarrow F'=15/2$ and $F=11/2\rightarrow F'=13/2$ transitions and return population to the $F=15/2$ level of the $\mathrm{a^5F_5}$ state. 
    }
    \label{fig:depumping-and-repumping}
\end{figure}

Rather, laser cooling is performed on atoms in a metastable electronic configuration, $\mathrm{3d^3(^4F)4s}$, specifically on the $\mathrm{a ^5F_5}$ term, to which we refer as ``the laser-cooling state.''  As demonstrated in Ref.~\cite{eustice_magneto-optical_2025}, the transition $\mathrm{3d^3(^4F)4s}$ $\mathrm{a^5F_5\rightarrow 3d^3(^4F)4p}$ $\mathrm{y^5G_6^o}$, at an optical wavelength of \SI{498}{nm} and a natural linewidth around $\gamma_\mathrm{498nm} = 2 \pi \times \qty{10.8}{\MHz}$, is a nearly closed transition.

Because the laser-cooling state lies at an energy of $\Delta E = h c \times \qty{6843}{\cm^{-1}}$ above the ground state, corresponding to $T = \Delta E/k_B \sim \qty{10 000}{\kelvin}$, a sublimated atomic beam of Ti contains only a very small population of atoms in the laser-cooling state.  To enrich the population in this state, we optically pump atoms from the $\mathrm{a ^3F_4}$ ground term, driving the transition $\mathrm{3d^24s^2}$ $\mathrm{a^3F_4\rightarrow 3d^2(^3P)4s4p(^3P^o)}$  $\mathrm{y^5D_4^o}$ at an optical wavelength of \qty{391}{\nm}.

In each of the previously trapped isotopes of titanium, $^{46}$Ti, $^{48}$Ti, and $^{50}$Ti, the nuclear spin is zero ($I=0$), meaning these isotopes lack hyperfine structure~\cite{eustice_magneto-optical_2025}.  The absence of hyperfine structure simplifies optical manipulation of the atom in that the entire population in each of the low-energy levels ($\mathrm{a ^3F_4}$ and $\mathrm{a ^5F_5}$) can be resonantly addressed by a single-tone light field.

In contrast, $^{47}$Ti and $^{49}$Ti have non-zero nuclear spins of $I=5/2$ and $I=7/2$, respectively, introducing hyperfine structure to their atomic terms and splitting optical transitions into multiplets of lines; see Fig.~\ref{fig:atom-struc}(b,c). This hyperfine structure affects both the optical pumping and laser cooling of fermionic Ti isotopes.  The effect of hyperfine structure on optical pumping is that only the population in one hyperfine level may be pumped and the spontaneous emission to the $\mathrm{a ^5F_5}$ term populates several hyperfine levels of the laser-cooling state, altogether reducing the optical pumping efficiency (see Fig.\ \ref{fig:depumping-and-repumping}(a)). 
The effect of hyperfine structure on the laser-cooling transition is that, even while near-resonantly driving the stretched state transition $\ket{F = I+J} \rightarrow \ket{F^\prime = I + J^\prime}$, off-resonant scattering can depump the hyperfine spin of atoms in the laser-cooling state.  If this hyperfine-level depumping is not counteracted, typically by applying additional ``repumping'' tones that resonantly drive atoms back to the stretched hyperfine states, an atom will cease to scatter laser-cooling light and be lost (see Fig.~\ref{fig:depumping-and-repumping}(b)).

To perform optical pumping and laser cooling of the fermionic isotopes of Ti, the hyperfine structure of the upper and lower states of both transitions must be known. To date, knowledge of the of this hyperfine structure in $^{47}$Ti and $^{49}$Ti, is incomplete; see Table \ref{tab:hyperfine-const-wide}.  In the next two Sections, we describe how we have determined the hyperfine structure of the relevant terms, and of the transitions between them, through a two-pronged approach: First, we use advanced atomic-structure calculations to parlay the existing knowledge of the $^{47}$Ti and $^{49}$Ti nuclei to predict theoretically the hyperfine structure of several atomic levels in Ti (Sec.~\ref{sec:theory}).  Second, we perform optical spectroscopy on a Ti atomic beam to measure the resonant frequencies of several lines within both the optical pumping and laser cooling hyperfine multiplets (Sec.~\ref{sec:spectroscopy}).

\begin{table*}[t]
    \centering
    \fontsize{8.5}{10}\selectfont
    \setlength{\tabcolsep}{1.5pt}
    \begin{tabular}{c|ccc|ccc|ccc|ccc}
    \hline \hline
    & \multicolumn{3}{c|}{$A_{47}/h$ (\unit{MHz})} & \multicolumn{3}{c|}{$B_{47}/h$ (\unit{MHz})} & \multicolumn{3}{c|}{$A_{49}/h$ (\unit{MHz})} & \multicolumn{3}{c}{$B_{49}/h$ (\unit{MHz})} \\
     term  & prev. exp. & exp. & thr. & prev. exp. & exp. & thr. & prev. exp. & exp. & thr. & prev. exp. & exp. & thr. \\ \hline
     $\mathrm{a^3F_3}$  & -52.9047(7) & --- & \bf{-51(3)} & 28.082(9) & --- & \bf{27(4)} & -52.1830(7) & --- & \bf{-51(3)} & 23.03(1) & --- & \bf{22(3)} \\
    $\mathrm{a^3F_4}$  & -37.9918(6) & \bf{-38.1(0.2)} & \bf{-37(4)} & 37.681(10) & \bf{35.0(3.2)} & \bf{36(5)} & -38.0042(5) & \bf{-37.9(0.2)} & \bf{-37(4)} & 30.842(13) & \bf{27.5(4.5)} & \bf{30(4)} \\
    $\mathrm{a^5F_5}$  & -74.8450(25) & \bf{-75.3(0.3)} & \bf{-69(10)} & -28.674(53) & \bf{-30.2(6.5)} & \bf{-27(4)} & -74.866(23) & \bf{-74.5(0.3)} & \bf{-69(10)} & -23.484(52) & \bf{-29.2(8.9)} & \bf{-22(3)}\\
    $\mathrm{z^5G_6^o}$ & --- & --- & \bf{-85(9)} & --- & --- & \bf{77(5)} & --- & --- & \bf{-86(9)} & --- & --- & \bf{63(4)} \\
    $\mathrm{y^5D_4^o}$ & --- & \bf{-96.6(0.2)} & \bf{-88(11)} & --- & \bf{-0.24(3.4)} & \bf{4(13)} & --- & \bf{-96.4(0.2)} & \bf{-88(11)} & --- & \bf{-3.3(4.4)} & \bf{-3(11)} \\
    $\mathrm{y^5G_6^o}$ & --- & \bf{-14.2(0.3)} & \bf{-12(3)} & --- & \bf{-1.1(6.9)} & \bf{-3(5)} & --- & \bf{-13.6(0.3)} & \bf{-12(3)} & --- & \bf{-5.0(10.9)} & \bf{-2(4)} \\\hline\hline
%
    \end{tabular}
    \caption{Hyperfine $A$ and $B$ constants of several terms for the $^{47}$Ti and $^{49}$Ti isotopes.  For each constant we present three values: previous experimental measurements reported in Refs.~\cite{channappa_hyperfine_structure_measurements_lowlying_multiplets_47ti_49ti_59co_105pd_1965,aydin_sternheimer_free_determination_the47ti_nuclear_quadrupole_moment_hyperfine_structure_measurements_1990,rumble_table_nodate} (prev.~exp.), results from fluorescence spectroscopy reported in this paper (exp.), and theoretical predictions derived in this paper by combining known nuclear moments with advanced atomic-structure calculations (thr.). All measurements from this paper are bolded.}
    \label{tab:hyperfine-const-wide}
\end{table*}

\section{Calculation of Ti HFS}
\label{sec:theory}

We write the hyperfine interaction Hamiltonian as
\begin{equation}
    H_{\mathrm{HFI}}=\sum_{k}\mathcal{M}^{(k)} \mathcal{T}^{(k)},
\end{equation}
where $\mathcal{M}^{(k)}$ are multipole nuclear moments of rank $k$ and $\mathcal{T}^{(k)}$ are electronic coupling operators of the same rank.  Each term with different $k$ contributes to the energy of a specific hyperfine level as follows:
\begin{equation}
    E^{(k)}_{\mathrm{HFI}}=\left\langle \gamma(IJ),F, M_F \left|\mathcal{M}^{(k)} \mathcal{T}^{(k)}\right|\gamma(IJ),F, M_F\right\rangle,
\label{A_k}
\end{equation}
where the nuclear and electronic angular momenta $I$ and $J$ are coupled to produce a state of definite total momentum $F$ and its projection $M_F$, and $\gamma$ encapsulates all other atomic quantum numbers. Using the Wigner-Eckart theorem, Eq.~\ref{A_k} may be written as
\begin{equation}
    \begin{aligned}
E^{(k)}_{\mathrm{HFI}} =  (-1)^{I+J+F}& \left\{\begin{array}{lll}
I & I & k \\
J & J & F
\end{array}\right\} \\
\times &\left\langle I\left\|\mathcal{M}^{(k)}\right\| I\right\rangle\left\langle \gamma, J\left\|\mathcal{T}^{(k)}\right\| \gamma, J\right\rangle
\end{aligned}
\end{equation}

In the present work, we restrict the hyperfine-interaction Hamiltonian $H_{\mathrm{HFI}}$ to contributions from the nuclear magnetic dipole moment ($k=1$) and the electric quadrupole moment ($k=2$). The corresponding reduced matrix elements for these nuclear moments are
\begin{equation}
\begin{aligned}
    \left\langle I\left\|\mathcal{M}^{(1)}\right\| I\right\rangle=&\mu_I\sqrt{\frac{(2 I+1)(I+1)}{I}} ,\\
    \left\langle I\left\|\mathcal{M}^{(2)}\right\| I\right\rangle=&\frac{Q}{2} \sqrt{\frac{(2 I+3)(2 I+1)(I+1)}{I(2 I-1)}}
\end{aligned}
\end{equation}
where the nuclear magnetic dipole moment is defined as $\mu_I=\langle I. M_I=I|\mathcal{M}^{(1)}|I,M_I=I\rangle$ and the nuclear electric quadrupole moment is defined as $Q=\langle I ,M_I=I|\mathcal{M}^{(2)}|I,M_I=I\rangle$.  Accepted measured values of both nuclear parameters are tabulated in Refs.\ \cite{Stone_m_moments, Stone_q_moments}.  Relevant to our work, for $^{47}$Ti, $\mu_I = -0.7865(4)\, \mu_N$ and $Q = \qty{0.302(10)}{\barn}$, while for $^{49}$Ti, $\mu_I = -1.10370(14) \, \mu_N$ and $Q = \qty{0.247(11)}{\barn}$; here $\mu_N$ is the nuclear magneton.  

When combined with matrix elements of the term-specific electronic coupling operators, these known nuclear moments determine the hyperfine constants $A$ and $B$, for each isotope and atomic term, in the the conventional form~\cite{johnson_atomic_structure_theory_lectures_atomic_physics_2007}:
\begin{equation}
    \begin{aligned}
A & =\frac{\mu_I}{I}\frac{\left\langle\gamma, J\left\|\mathcal{T}^{(1)}\right\| \gamma, J\right\rangle}{\sqrt{J(J+1)(2 J+1)}}, \\
B & =2Q\sqrt{\frac{J(2 J-1)}{(2 J+3)(2 J+1)(J+1)}}\left\langle\gamma , J\left\|\mathcal{T}^{(2)}\right\| \gamma, J\right\rangle
\end{aligned}
\end{equation}
These $A$ and $B$ coefficients then determine the hyperfine-state energies at each hyperfine spin quantum number $F$, at zero magnetic field, as
\begin{equation}
    E_\mathrm{HFI} = \frac{K}{2} A + \frac{1}{2} \frac{3 K (K+1) - 4 J (J+1) I (I+1)}{2 I (2I-1) 2 J (2J-1)} B
    \label{eq:hyperfine}
\end{equation}
with $K = F(F+1) - I(I+1) - J(J+1)$.

What remains is to calculate the matrix elements of the electronic coupling operators.  For this, we perform theoretical calculations of the atomic structure of Ti using the powerful and precise CI + all-order method, which combines linearized couple-cluster (CC) and configuration interaction approaches to calculate wavefunctions and energies of atomic states~\cite{cheung_pci_2025}.
In our previous work, we calculated the energies and wavefunctions for the 85 lowest levels of neutral Ti, finding differences between the calculated energies and experimentally measured energies of only 0.1-2.8\%, indicating good agreement~\cite{eustice_optical_2023}. In this work, we use the calculated wavefunctions to obtain theoretical values for the hyperfine constants of the levels relevant to laser cooling.

When calculating hyperfine matrix elements, we include corrections to the $\mathcal{T}^{(k)}$ operator beyond random phase approximation (RPA), such as core-Brueckner ($\sigma$), structural radiation (SR), two-particle (2P) and normalization (Norm) ~\cite{dzuba_using_effective_operators_calculating_hyperfine_structure_atoms_1998,porsev_calculation_hyperfine_structure_constants_ytterbium_1999,porsev_electricdipole_amplitudes_lifetimes_polarizabilities_lowlying_levels_atomic_ytterbium_1999}. While these corrections are generally small and tend to cancel for transition properties, they should be included for hyperfine constants. The final results are presented in Table~\ref{tab:hyperfine-const-wide}. There are two main sources of the uncertainties of the theoretical computations: (1) uncertainty associated with the treatment of core correlations  $\Delta_{\mathrm{c}}$ and (2) uncertainty in the computation of the corrections to the hyperfine operator $\Delta_{\mathrm{o}}$. The value of  $\Delta_{\mathrm{c}}$ is estimated as the difference between the results obtained with the more accurate CI+all-order method and the CI+MBPT method, where the effective Hamiltonian is constructed using second-order many-body perturbation theory (MBPT). The value of $\Delta_{\mathrm{o}}$ is estimated assuming 50\% uncertainty in each of four correction beyond RPA added in quadrature. The resulting total theory uncertainty estimate  is calculated as
\begin{equation}
\Delta = \sqrt{\Delta_{\mathrm{c}}^2+ \Delta_{\mathrm{o}}^2}.
\end{equation}


\section{Hyperfine spectroscopy}
\label{sec:spectroscopy}

\begin{figure}
    \centering
    \includegraphics[]{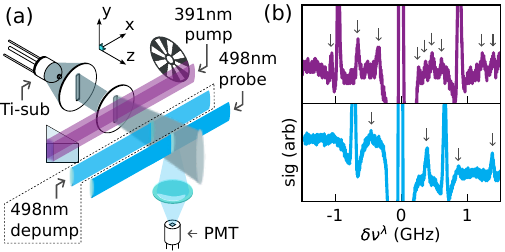}
    \caption{(a) Diagram of the atomic beam spectroscopy setup. Sublimated Ti atoms are collimated by two slits to form an atomic beam traveling along the $z$ direction. The atomic beam first passes through a pair of counter-propagating \qty{391}{\nm} wavelength optical pumping laser beams before entering the \qty{498}{\nm} wavelength probe laser beam, at which point the atomic fluorescence is imaged onto a photomultiplier tube (PMT). An optical chopper in the path of the \SI{391}{\nm} wavelength light enables lock-in amplification of the fluorescence signal.
    In a second configuration, indicated by the dotted outline, an additional \SI{498}{\nm} wavelength depumping beam is introduced before the probe beam in order to measure splittings between $F$ levels in the $\mathrm{y^5G_6^o}$ term.
    (b) Spectra observed through one-color fluorescence spectroscopy. The frequencies of the \qty{391}{nm} wavelength optical pumping (top) or the \qty{498}{nm} wavelength laser cooling transition (bottom) are scanned broadly across the resonances of the 5 different isotopes. Frequencies, $\delta\nu^\lambda$, are given relative to the corresponding resonance frequency in $^{48}$Ti. The probe light is frequency modulated at frequency $f_\mathrm{mod}$, and then the PMT signal is demodulated at $2 f_\mathrm{mod}$ to observe resonance lines with reduced noise.  These spectra are dominated by strong signals from each of the three bosonic Ti isotopes.  A few additional features (indicated with arrows) are identified, arising from the fermionic isotopes.  However, these additional features are insufficient to determine the full hyperfine structure of the relevant transitions, necessitating a two-color spectroscopic method. 
    }
    \label{fig:spectroscopy-schematic}
\end{figure}

To complement these theoretical calculations, we measured the hyperfine structure of the optical pumping and laser cooling transitions of both $^{47}$Ti and $^{49}$Ti by two-color fluorescence spectroscopy of a Ti atomic beam (see Fig.~\ref{fig:spectroscopy-schematic}).  The atomic beam was generated using a Ti-sublimation vacuum pump as a source of sublimated atoms~\cite{schrott_atomic_2024}.  The beam was collimated along one axis ($x$) by a pair of $\SI{5}{\mm}\times\qty{20}{\mm}$ slits separated by \SI{10}{\cm}, resulting in a beam that diverges with a half-angle of \SI{0.032}{rad} in the $x$ direction. The flux density of $\mathrm{a^3F_4}$ ($\mathrm{a^5F_5}$) atoms at the location of the probe laser is estimated to be $\sim$\SI{1e10}{\s^{-1}\cm^{-2}} (\SI{1e7}{\s^{-1}\cm^{-2}}), and the mean longitudinal velocity of the beam is expected to be $\bar{v}_L=\SI{780}{\meter/\second}$. 

The collimated beam was illuminated with coherent light beams with frequencies tuned across either the optical pumping (\qty{391}{\nm} wavelength) or laser cooling (\qty{498}{\nm} wavelength) transitions.  These light beams were directed transverse to the emission of the atom beam and along the narrowly collimated direction ($x$) of the atomic beam, and were aligned to be closely parallel to one another. With the atomic velocity distribution along $x$ having an estimated full width at half maximum (FWHM) of \SI{50}{\m/\s}, we expect the atomic fluorescence spectrum from these transverse probe beams to be Doppler broadened (FWHM) to \SI{130}{\MHz} for the optical pumping transition, and \SI{100}{\MHz} for the laser cooling transition.  Fluoresced light was imaged (numerical aperture of 0.3, magnification of 1) onto a bialkali photomultiplier tube.

 \begin{figure*}[ht]
    \centering
    \includegraphics[]{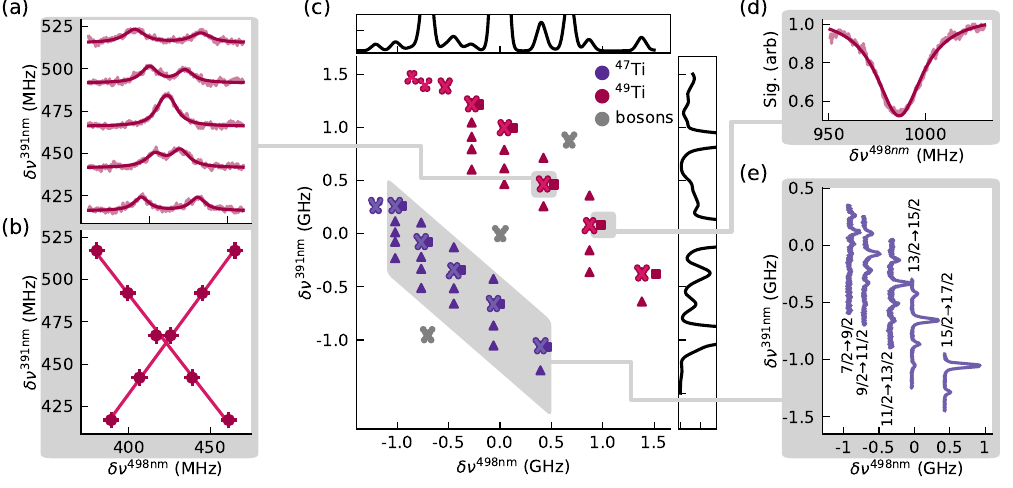}
    \caption{(a,b) Simultaneous determination of optical pumping and laser cooling resonances using the ``X marks the spot" scheme. When tuned near a line in the optical pumping multiplet, the double-passed \SI{391}{\nm} wavelength light pumps atoms with opposite transverse velocities into the $\mathrm{a^5F_5}$ manifold, causing two peaks to be observed for a given line in the \SI{498}{\nm} wavelength fluorescence signal. By stepping the optical pumping light frequency and scanning the \SI{498}{\nm} wavelength light, the spectra shown in (a) are obtained, with the optical-pumping light frequency indicated by the vertical offset. Data and fits are shown in light and dark pink respectively. The fit locations of the peaks are shown in (b), with the crossing point indicating the zero-velocity frequency of both the optical pumping and laser cooling resonances.
    (c) By repeating this procedure for a variety of optical pumping and laser cooling transitions on each isotope, more of the overall structure is measured. Each ``X'' on the plot is a dataset like the one shown in part (a). The frequencies of the optical pumping resonances ($\delta\nu^\mathrm{391nm}$) and the laser cooling resonances ($\delta\nu^\mathrm{498nm}$) are again given relative to the resonances in $^{48}$Ti. At top (right), the theoretical one-color fluorescence spectra for the laser cooling (optical pumping) transition multiplet is plotted, demonstrating the strength of the two-color scheme for resolving obscured lines. The strong lines from the bosonic isotopes are above the vertical scale of the plot. (d) Three-color spectroscopy. Weak $F\rightarrow F$ laser cooling resonances (shown as squares in (c)) were observed by tuning the optical pumping and laser cooling light to the center of an ``X'' and adding an additional \qty{498}{nm} wavelength beam that depumps atoms from the hyperfine level addressed by the fluorescence beam. The figure shows the depletion of fluorescence on the $F\rightarrow F+1$ transitions as the $F\rightarrow F$ depumping beam is scanned through resonance. (e) Additional optical pumping resonances were observed by tuning the probe beam to a laser cooling resonance and broadly scanning the optical pumping light frequency. Multiple optical pumping lines can optically pump into a given $\mathrm{a^5F_5}$ level. When the optical pumping light is scanned through one of these lines, a peak is observed in the probe fluorescence. Each trace is labeled by the laser-cooling line addressed by the probe beam, and the horizontal offset of each trace shows the probe frequency. The resonances observed this way are indicated by triangles in (c).}
   \label{fig:treasure-map}
\end{figure*}

Figure~\ref{fig:spectroscopy-schematic}(b) shows fluorescence spectra obtained by one-color probing, i.e.\ applying only either the optical pumping light or the laser cooling light.  For these data, the central optical frequency was slowly scanned several GHz across the relevant atomic transition.  The light was additionally dithered around its central frequency and the fluorescence signal demodulated at twice the modulation frequency, leading to the Ricker wavelet-like lineshapes seen in the spectra.  On both transitions, the spectrum is dominated by the strong resonances of the $I=0$ isotopes, $^{46}$Ti, $^{48}$Ti, and $^{50}$Ti.  For these, the lack of hyperfine structure concentrates the transition line strength at a single frequency.  A few weaker resonance signals are observed from the $I\neq 0$ isotopes, $^{47}$Ti and $^{49}$Ti.  However their full spectrum, comprised of numerous lines connecting the different lower- to upper-level hyperfine states, is unresolved.  Indeed, as shown by the theoretical curves plotted along with Fig.~\ref{fig:treasure-map}(c), several of their optical transitions lie in close proximity to and are then obscured by the strong $I=0$ isotope resonances.

To resolve the transition multiplets of the fermionic isotopes, we employed a two-color scheme, pumping the atomic beam with optical pumping light and then measuring fluorescence with laser cooling light. In this method, the frequency of the \qty{391}{\nm} wavelength light is tuned to drive a line in the optical pumping multiplet of a chosen fermionic isotope.  This optical pumping increases the population in up to three hyperfine levels of the $\mathrm{a^5F_5}$ manifold.  Increased fluorescence of the \qty{498}{\nm} wavelength light detects these enriched populations, allowing us to identify coincidentally both the optical pumping and the laser cooling resonances.  We detected this increased fluorescence by intensity-modulating the optical pumping light using a chopper wheel.  We used lock-in detection to measure the contribution of optically pumped atoms, even against the potentially large background signal from the strong (but not pumped) bosonic isotope lines.

The pump-probe scheme also reduced the Doppler width of the pumping-enhanced, 498-nm-wavelength resonances.  The transverse velocity distribution of the optically pumped population is determined by the power-broadened linewidth of the optical pumping transition. In our experiment, the intensity of the optical pumping light was about 12 times the saturation intensity for this transition.  The atomic response was thereby power broadened from the natural linewidth of \SI{414}{\kHz} to \SI{1.5}{\MHz}, leading to an optically pumped beam with a Lorentzian transverse velocity distribution of FWHM $\sim$\SI{0.6}{\m/\s}, much narrower than the transverse velocity of the collimated sublimated atomic beam itself.  This narrowing of the velocity distribution of the optically pumped beam led to the inhomogeneous Doppler broadening of the laser-cooling transition being negligible in comparison to the  \SI{10.8}{\MHz} homogeneous linewidth.

Special care was taken to ensure the observed resonances correspond to atoms with zero transverse velocity. A prism retro-reflector was used to send counter-propagating passes of optical pumping light through the atomic beam. When detuned slightly from the zero-velocity resonance of an optical pumping line, the counter-propagating beams address atoms at velocities $\pm v_x$, resulting in a double-peak feature in the 498-nm-wavelength fluorescence spectrum (Fig.\ \ref{fig:treasure-map}(a)).  As the frequency of the optical pumping beam is tuned, these two peaks cross, marking the frequencies of the zero-velocity resonances for both the optical pumping and laser cooling transitions (Fig.~\ref{fig:treasure-map}(b)).

We used this ``X marks the spot'' procedure to measure all of the strong $F\rightarrow F+1$ transitions in the laser-cooling multiplets of both isotopes. For each laser cooling line, a convenient, strong optical pumping line was chosen from the $F\rightarrow F$,  $F\rightarrow F-1$, or $F\rightarrow F+1$ series, with the resulting X that indicates the zero velocity resonance shown in Fig.~\ref{fig:treasure-map}(c). Once the frequency of a laser cooling line was established, more optical pumping lines were observed by tuning the 498-nm-wavelength light to the zero-velocity resonance and performing a broad frequency scan of the 319-nm-wavelength light. The locations of optical pumping lines found this way are shown with single triangular markers in Fig~\ref{fig:treasure-map}(c) with representative spectra shown in Fig~\ref{fig:treasure-map}(e).

This two-color fluorescence method was sufficient to measure all the $F \rightarrow F+1$ transitions of the laser-cooling line, transitions on which many photons are fluoresced before the atom is depumped to an undriven hyperfine state.  However, this sequence of resonances alone is insufficient to accurately determine the hyperfine structure of each of the two states connected by these transitions.  Therefore, we performed an additional \emph{three}-color spectroscopy:  After selectively optically pumping fermionic-isotope atoms, and before probing their fluorescence on a bright $F \rightarrow F+1$ laser-cooling transition, we drove them with an additional hyperfine-depumping tone.  When this hyperfine-depumping tone was resonant with an $F\rightarrow F$ resonance on the laser-cooling line, we observed a reduction in the probe fluorescence. Laser-cooling lines measured this way are shown with square markers in Fig.~\ref{fig:treasure-map}(c)  with a representative signal shown in Fig~\ref{fig:treasure-map}(d).

The fermionic-isotope optical pumping and laser cooling transition frequencies measured by these methods are tabulated in Appendix~\ref{sec:table-of-measurments}, with all frequencies referenced to the respective $^{48}\mathrm{Ti}$ transition.  We fitted all measured resonances to the hyperfine multiplet predicted by application of Eq.~\ref{eq:hyperfine} to the upper and lower term of the transition, and offset by an overall isotope shift.  Least-squares fitting of this model to the data yielded the hyperfine $A$ and $B$ coefficients presented in Table~\ref{tab:hyperfine-const-wide}. We note the good agreement of theory, our experiments, and previous experiments, with all coefficients agreeing within estimated uncertainties. Appendix~\ref{sec:king-plot-analysis} presents a King plot analysis of the measured isotope shifts.

\section{Magneto-optical trapping of $^{47}$Ti and $^{49}$Ti}
\label{sec:mots}

\begin{figure}[t]
    \centering
    \includegraphics[]{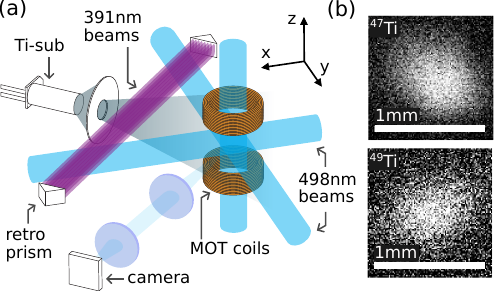}
    \caption{(a) Apparatus used to produce magneto-optical traps of $^{47}$Ti and $^{49}$Ti. The direct flux of Ti atoms from a Ti sublimation pump is sent through several passes of OP light before encountering a standard six-beam magneto-optical trap. Fluorescence and absorption images are taken through a 4-$f$ imaging system as depicted (not to scale). (b) Fluorescence images of trapped clouds of $^{47}$Ti and $^{49}$Ti atoms.}
    \label{fig:mot-diagram}
\end{figure}

Using the findings of the previous section, we turn to producing magneto-optical traps of $^{47}$Ti and $^{49}$Ti. The laser-cooling apparatus, shown in Fig.~\ref{fig:mot-diagram}(a), is the same as the one used in Ref.~\cite{eustice_magneto-optical_2025}.  In this apparatus, a sublimated atomic beam of Ti is optically pumped and then sent directly into a magneto-optical trapping region.  The small fraction of atoms with forward velocities below the magneto-optical trap's capture velocity ($\sim$\qty{50}{\m/\s}) can be cooled and trapped. 

Laser cooling is performed using light that is red-detuned by one to four linewidths from the stretched $\ket{F = I+J} \rightarrow \ket{F^\prime = I + J^\prime}$ transition within the laser-cooling multiplet.  For the bosonic isotopes of Ti, for which $I=0$, applying just this single tone of laser-cooling light is sufficient to form a magneto-optical trap. For the fermionic isotopes, to counteract the leakage of atoms out of the $\ket{F = I+J}$ hyperfine level, we apply additional hyperfine-repumping tones of light.  These tones are resonant with the $\ket{F=I+J-1} \rightarrow \ket{F'=F+1}$ and $\ket{F=I+J-2} \rightarrow \ket{F'=F+1}$ transitions, to which we refer as the repump one (RP1) and repump two (RP2) transitions, respectively (see Fig.~\ref{fig:depumping-and-repumping}(b)).   Experimentally, these tones are generated starting with a single-frequency light beam and using two acousto-optical modulators, each in a double-pass configuration, to shift the frequencies of portions of the beam by the requisite amounts.  This setup delivers a total of \qty{90}{\mW} of power in the laser-cooling tone and \qty{20}{\mW} in each of the repuming tones.  The magneto-optical trap is driven with six individually adjustable beams, with the combined intensity from all beams at the trap center being \SI{31}{\mW/\cm^2} for the laser-cooling tone and \SI{5.6}{\mW/\cm^2} for each of the repumping tones.

 \begin{figure*}[ht]
    \centering
    \includegraphics[]{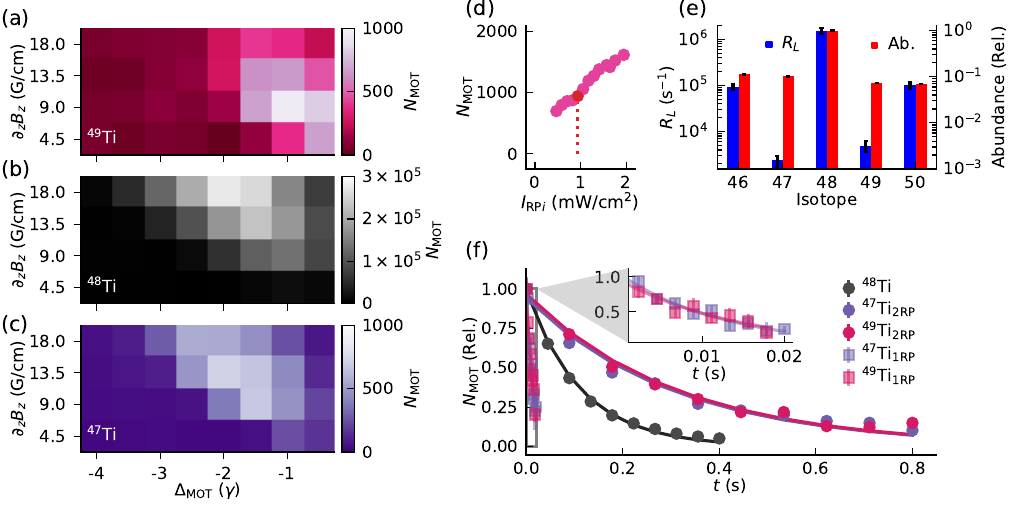}
    \caption{(a,b,c) Optimization of the number of $^{49}$Ti (a), $^{48}$Ti (b), and $^{47}$Ti (c) atoms in the magneto-optical trap as a function of the detuning of the primary laser cooling light frequency, $\Delta_{\mathrm{MOT}}$, and magnetic field gradient, $\partial_zB_z$. (d) The number of $^{49}$Ti atoms trapped in the system as the optical intensity in each repump beam is tuned. The dotted line indicates the intensity at which the rest of the data in the paper was taken. (e) Measured loading rates, $R_L$ (blue), and natural abundances (red) for the five isotopes of Ti. (f) Decay of the number of $^{48}$Ti, $^{47}$Ti, and $^{49}$Ti atoms in the trap after shuttering the optical pumping laser are shown in gray, purple and pink respectively. For the fermionic isotopes, magneto-optical traps operating with both one (light square markers) and two (darker circular markers) repump tones are shown, with the inset detailing the decay of the one repump traps. Solid lines show fits of the data to exponential decay, from which atom number lifetime is determined.}
    \label{fig:mot-plots}
\end{figure*}

Magneto-optical traps were realized within our setup for each of the five naturally abundant isotopes of Ti.  The magneto-optical traps of the fermionic $I\neq 0$ isotopes $^{47}$Ti and $^{49}$Ti differ from the previously described traps of the bosonic $I=0$ isotopes, $^{46}$Ti, $^{48}$Ti and $^{50}$Ti, in three significant ways.

First, in our setup we find that we can only load magneto-optical traps of the fermionic isotopes at low gradients of the spherical-quadrupole magnetic field.  Figures \ref{fig:mot-plots}(a-c) compare the steady-state number of magneto-optically trapped atoms of three isotopes of Ti.  To make this comparison, we load atoms into the trap, wait for the atom number to reach a steady state, switch the trap to a setting of laser power, detuning, and field gradient that is common for all measurements, and measure the atom number by imaging the fluorescence of laser-cooling light. We observe at most 731(190), \num{1.4(2)e5}, and 1142(240) atoms of $^{47}$Ti, $^{48}$Ti, and $^{49}$Ti respectively. In $^{48}$Ti, the atom number was determined by performing absorption imaging of the atoms shortly after switching off the trapping light and magnetic field. For the fermions, the number of trapped atoms was deduced from fluorescence data using an estimate of the atomic scattering rate as discussed below.  As reported previously \cite{eustice_magneto-optical_2025}, traps for the bosonic isotopes, such as $^{48}$Ti, increase monotonically in atom number as one increases simultaneously the detuning $|\Delta_\mathrm{MOT}|$ of the laser-cooling light from resonance, the magnetic field gradient $\partial_z B_z$, and the power of the laser-cooling light.  We understand that increasing these three trapping parameters increases the capture velocity and the capture area of the magneto-optical trap, allowing it to capture a larger fraction of atoms from the uncollimated and broadly velocity-distributed emission of the sublimation source.

By contrast, magneto-optical traps for the two fermionic isotopes reach their maximum atom numbers at a lower, bounded value of both $|\Delta_\mathrm{MOT}|$ and $\partial_z B_z$.  This bound may be explained by the effect of magnetic fields on the hyperfine structure of the excited state of the laser-cooling transition.  At large field gradients, the magnetic field at the periphery of the magneto-optical trapping region, i.e.\ about \qty{1}{\cm} from the trap center, causes the excited state hyperfine levels to become mixtures of states with different total angular momentum $F^\prime$.  Under such $F^\prime$ mixing, repumping becomes less effective and, indeed, can actively depump atoms to states in the $\mathrm{a ^5F_5}$ term with lower values of $F$ that are not addressed by our two repump tones.  Such depumping would reduce the loading rate of the magneto-optical trap. For the remaining data in this section, all magneto-optical traps were produced with detunings of $\Delta_\mathrm{MOT}=-1\gamma_\mathrm{498nm}$ and magnetic field gradients of $\partial_zB_z=$\qty{9}{G/\cm} to allow for a clearer comparison.

Second, we find that the lifetime of fermionic atoms in the magneto-optical trap is very low in the absence of repumping tones, but is restored by the addition of repumping tones.  We measure the trap lifetime by first loading the magneto-optical trap to a steady-state atom number and then suddenly extinguishing the optical pumping light.  This extinction reduces the loading rate of the trap, causing the magneto-optical trap to decay to a much lower atom number; see Fig.~\ref{fig:mot-plots}(f).  We fit this decay in the atom number vs.\ time to an exponential decay model to determine the magneto-optical trapping lifetime.

With the addition of the two resonant repump tones, the fermionic Ti magneto-optical trapping lifetimes are long: 330(15) ms and 310(8) ms for $^{47}$Ti and $^{49}$Ti, respectively. In contrast, extinguishing one of the repump tones (RP2) reduces the magneto-optical trap lifetime to \qty{13(1)}{ms} and \qty{15(1)}{ms}.  The trap lifetime with both repump tones extinguished was too short for us to measure.

To understand the observed lifetimes of the fermionic magneto-optical traps, we performed simulations of Ti atoms prepared in the stretched hyperfine state of the laser-cooling term and driven by light of isotropic polarization, matching the intensities at the center of our magneto-optical trap, under each of three conditions: (1) only laser-cooling light (detuned by one full linewidth, $\gamma_\mathrm{498nm}$, below the stretched hyperfine transition), (2) laser-cooling light and the resonant RP1 tone, and (3) laser-cooling light and both the resonant RP1 and RP2 tones.  Considering an atom to be lost once it occupies a hyperfine state no longer driven with near-resonant light, we find that a $^{47}$Ti ($^{49}$Ti) atom will be lost after an average of 1300(325) (1800(450)) photon scatterings under condition (1), \qty{9.6(6)e4} (\qty{9.9(13)e4}) photon scatterings under condition (2), and \qty{4.3(9)e6} (\qty{4.4(10)e6}) photon scatterings under condition (3). We refer to these quantities henceforth as ``photon budgets". While the lifetimes of the magneto-optically trapped fermions with both repumps applied may be limited by a number of processes, such as decay of the $\mathrm{y^5G^o_6}$ state to terms other than the $\mathrm{a^5F_5}$ or two-body loss, the lifetime when only one repump is applied can be assumed to be limited by depumping among hyperfine levels alone. Based on the observed lifetimes of the one-repump magneto-optical traps and the calculated photon budgets, we deduce single atom scattering rates of $\Gamma_{47}=$\qty{7.4(7)e6}{s^{-1}} and $\Gamma_{49}=$\qty{6.8(10)e6}{s^{-1}} for the fermions, in good agreements with measurements of the scattering rate of the trapped $^{48}$Ti atoms, which were done by combining measured fluorescence of the trapped atoms with measurements of the atom number using absorption imaging ($\Gamma_{48}=$\qty{7.8(9)e6}{s^{-1}}). We note that the intensity of our repump beams is high enough so that atoms spend little time in the $|F=I+J-1\rangle$ and $|F=I+J-2\rangle$ states, and the total scattering rate is expected to be limited by the excitation rate on the stretched transition. Given the scattering rates inferred from the single-repump lifetime, for a two-repump magneto-optical trap limited by hyperfine state depumping, one would expect lifetimes of $\sim$\qty{600}{ms} for $^{47}$Ti and $^{49}$Ti, considerably longer than the lifetimes we observe for both MOTs. We therefore conclude the lifetimes of the two-repump fermion magneto-optical traps are not primarily limited by hyperfine depumping, and would not significantly benefit from the addition of a third repumping tone.

We note that the lifetimes of our fermionic-atom traps are somewhat longer than the $\tau_{48}=$\qty{130(11)}{\ms} lifetime observed for bosonic $^{48}$Ti (with the same \qty{90}{\mW} of laser-cooling light). Considering that we deduce comparable atomic scattering rates for all three trapped isotopes, we conclude that an additional loss mechanism is present in the case of $^{48}$Ti. To clarify the sources of loss in the 2-repump fermion and the $^{48}$Ti magneto-optical traps, we investigate the leakage branching of the $\mathrm{y^5G^o_6}$ state to dark terms other than the $\mathrm{a^5F_5}$ state. Taking the measured scattering rate and lifetime of the trapped $^{48}$Ti atoms in Fig.~\ref{fig:mot-plots}(f), and considering the case that leakage out of the cooling transition is the only loss mechanism, we constrain the leakage branching ratio out of the $\mathrm{y^5G^o_6}$ term to be $\alpha\leq 1/\tau_{48}\Gamma_{48} =$\qty{9.9(2.1)e-7}.

Given the evidence of an additional loss mechanism in the $^{48}$Ti trap, we examined this bound further by measuring per-atom loss rates of $^{48}$Ti magneto-optical traps at lower powers of laser cooling light. We find evidence that these loss rates vary nonlinearly with the power in the laser beams and also with detuning. From measurements of scattering rates and trap lifetimes at low optical powers, again ascribing all loss to leakage, we obtain a stricter upper bound of $\alpha \leq 1/(\Gamma_{48} \tau_{48}) = 4.5(1) \times 10^{-7}$). This measured value agrees reasonably well with the predicted branching ratio of $\alpha = 1.1 \times 10^{-6}$ determined from atomic-structure calculations in Ref.~\cite{eustice_optical_2023}. Combining the photon budget implied by the leakage branching ratio (\num{2.2e6}) with the scattering rates determined for the fermionic isotopes, we predict lifetimes of \qty{300}{\ms} and \qty{320}{\ms} for these traps, in fairly good agreement with the observed values. This suggests that the lifetimes of the fermion magneto-optical traps are limited by the same process that limits the lifetime of the low-power $^{48}$Ti trap. Seeing as the density of the trapped fermionic atoms and the trapped $^{48}$Ti atoms differ by over two orders of magnitude, we conclude that the two-repump fermion magneto-optical traps are likely limited by branching from the $\mathrm{y^5G^o_6}$ term to terms besides the $\mathrm{a^5F_5}$ term. The additional sources of loss in the high-power $^{48}$Ti magneto-optical traps have not yet been investigated, but may suggest that light induced losses are higher for traps produced at $\Delta_\mathrm{MOT}=-1\, \gamma_\mathrm{498nm}$ than were measured in traps produced with larger absolute detuning ($\Delta_\mathrm{MOT}=-5\, \gamma_\mathrm{498nm}$)\cite{eustice_magneto-optical_2025}.

Third, we find that the magneto-optical-trap loading rate is smaller for the fermionic isotopes than for the bosonic ones.   The loading rate, reported in Fig.~\ref{fig:mot-plots}(e), is determined by switching on the trap for a variable time before counting the number of trapped atoms and extracting the loading rate from the initial linear rise in this number. We find loading rates of \qty{9.4(1.5)e4}{\s^{-1}}, \qty{2.4(6)e3}{\s^{-1}}, \qty{1.5(3)e6}{\s^{-1}}, \qty{6.0(1.4)e3}{\s^{-1}}, and \qty{1.0(2)e5}{\s^{-1}} for $^{46}$Ti-$^{50}$Ti. Normalizing for isotopic abundance, the loading rates of the fermions are significantly lower than those of the $^{48}$Ti trap, by factors of 65 for $^{47}$Ti and 19 for $^{49}$Ti. Part of this reduction is expected on account of the reduced optical-pumping efficiency. When driving the the $13/2\rightarrow13/2$ ($15/2\rightarrow15/2$) transition in the optical pumping multiplet in $^{47}$Ti ($^{49}$Ti), a fraction of 0.26 (0.22) of the $\mathrm{a^3F_4}$ population is addressed. From the upper $F'=13/2$ ($F'=15/2$) state of the optical pumping transition, the three highest $F$ levels of the $\mathrm{a^5F_5}$ level may be populated, each of which is addressed by a repump or laser cooling beam. Accounting for this effect, we still find lower loading rates in $^{47}$Ti and $^{49}$Ti by factors of 17 and 4. This may be explained in part by the low repump power in our system. Fig.~\ref{fig:mot-plots}(d) shows the number of atoms in the $^{49}$Ti trap as the intensity in each repump is increased. A factor of 1.7 increase in the $^{49}$Ti loading rate is observed as the repump intensity is increased by a factor of 2 (no change in the trap lifetime was observed as the repump intensity was scanned across this range). We suspect Zeeman broadening at the edges of the magneto-optical trap may hamper the repumping of the atoms, requiring power broadening to overcome. Higher repump powers in our system should be possible with straightforward improvements to the optical chain.

\section{Conclusion}
\label{sec:conclusion}

In conclusion, we present two essential steps to bringing the two naturally abundant, non-zero nuclear spin, fermionic isotopes of Ti into the ultracold regime.  First, we theoretically calculate and measure hyperfine structure in Ti.   Our theory and experiment results agree with previous experiments on the long-lived terms $\mathrm{a ^3F_4}$ and $\mathrm{a ^5F_5}$.  The theory and experiment results also agree with one another, giving confidence in the atomic structure calculations that underlie the theoretical predictions.  We note that the precision of our experimental measurements of the hyperfine splittings is limited by the linewidth of the optical transitions used to measure such splittings and by minute to minute drifts of our wavemeter with which the frequency splittings in the multiplet structure were measured.  These measurements can be refined in the future using better frequency references and by performing spectroscopy on laser-cooled atoms for which Doppler and transit-time broadening is reduced as compared to atomic beams.

Second, we report on laser cooling and trapping of the $I\neq0$ isotopes $^{47}$Ti and $^{49}$Ti.  The simple experimental setup used to achieve these results limits us to studying small laser-cooled gas samples, containing fewer than $10^4$ atoms.  Nevertheless, the setup allows us to identify several characteristics of the fermionic-isotope traps that differ from those of the bosonic isotopes of Ti.  Our results point to several areas for improvement, including the use of multi-tone optical pumping to transfer a larger population of atoms from the various $\mathrm{a ^3F_4}$ hyperfine states into the $\mathrm{a ^5F_5}$ laser cooling state, and higher power in the repumping tones. The fermionic isotopes may also benefit from a ``dark" magneto-optical trap configuration, where high optical power is used in the first repump, but low optical power is used in the second repump. Such a scheme would cause atoms to dwell for a significant amount of time in the $\ket{F=I=J-2}$ state, reducing the losses induced by the strong primary laser-cooling tone without hampering the loading rate of the trap~\cite{ketterle_high_densities_cold_atoms_dark_spontaneousforce_optical_trap_1993,townsend_highdensity_trapping_cesium_atoms_dark_magnetooptical_trap_1996}.  We note the small atom numbers precluded us from measuring the temperatures achieved in the fermionic-atom magneto-optical traps.  We expect that polarization gradient cooling of such gases produced temperatures in the range of tens of \unit{\micro\kelvin}, as was observed for the bosonic isotopes~\cite{eustice_magneto-optical_2025}. 

Adding ultracold gases of $^{47}$Ti and $^{49}$Ti to the family of atomic Fermi gases has the potential to open up several lines of scientific inquiry.  As discussed in Ref.~\cite{eustice_laser_2020}, the atomic structure of Ti results in a strongly anisotropic optical polarizability for ground state atoms, even far from optical resonances.  This feature implies that Ti atoms can be manipulated with highly coherent state-dependent forces and Raman transitions, with applications in evaporation~\cite{lopes_radiofrequency_evaporation_optical_dipole_trap_2021}, internal state control~\cite{lee_engineering_large_stark_shifts_control_individual_clock_state_qubits_2016}, and spin-orbit coupling~\cite{eustice_laser_2020}; having relevance to the production of large quantum gases, quantum information processing, and quantum simulation, respectively. Additionally, the low-but-nonzero magnetic moment of ground state Ti atoms minimizes the effect of long-range dipolar interactions while allowing $s$-wave contact interactions to be tuned by Feshbach resonances. These features position $^{47}$Ti and $^{49}$Ti as a complement to experiments on lanthanide and alkali atoms. 

\begin{acknowledgments}
This material is based upon work supported by the U.S. Department of Energy, Office of Science, National Quantum Information Science Research Centers, Quantum Systems Accelerator.  Additional support is acknowledged from the ONR (Grants No.~N00014-25-12105 and No.~N00014-22-1-2280), the ARO (Grants No.~W911NF2010266 and No.~W911NF2310244), the NSF (PHY-2012068, PHY-2309254, and the QLCI program through Grant No.~OMA-2016245), and the Heising-Simons Foundation (Award No.~2018-0904). R.D. acknowledges support from the ONR through the NDSEG Fellowship.

\end{acknowledgments}

\appendix

\section{Measured hyperfine and isotope shifts}
\label{sec:table-of-measurments}

After collecting spectra using the two- and three-color spectroscopy methods described in Sec.\ \ref{sec:spectroscopy}, we performed least-squares fitting to determine the frequency shift of the hyperfine and isotope-specific lines. Table~\ref{tab:isotope-shift-measurements} lists the results for the isotope shifts of both the bosonic and fermionic isotopes, while Table~\ref{tab:hyperfine-splittings} lists the fitted frequencies for all of the measured hyperfine transitions.

\begin{table}[h!]
    \centering
    \fontsize{8.5}{10}\selectfont
    \setlength{\tabcolsep}{1.5pt}
    \begin{tabular}{cccccc}
        \hline \hline
        $\lambda$ & $\delta\nu^\lambda_{46}$ & $\delta\nu^{\lambda,\mathrm{ctr}}_{47}$ & $\delta\nu^{\lambda, \mathrm{ctr}}_{49}$ & $\delta\nu^\lambda_{50}$ & Ref. \\
        
        (\si{\nm}) & (\si{\MHz}) & (\si{\MHz}) & (\si{\MHz}) & (\si{\MHz}) & \\
        \hline
        391 & \bf{-942.6(3.1)} & \bf{-447.8(3.1)} & \bf{474.2(3.1)} &  \bf{883.9(3.3)} &  \\ 
         & -950(2) & --- & --- &  -880(2) & \cite{eustice_magneto-optical_2025} \\ \hline
         498 & \bf{-708.3(3.0)} & \bf{-345.6(3.2)} & \bf{354.9(3.2)} & \bf{670.7(3.2)} &  \\
         & -703(2) & --- & --- &  673(2) & \cite{eustice_magneto-optical_2025} \\
         & -716.19(39) & --- & --- &  671.6(21) &  \cite{neely_isotope_2021} \\  \hline \hline
    \end{tabular}
    \caption{Isotope shift measurements of Ti transitions. Each transition is indexed by its optical wavelength, $\lambda$, in nm. All shifts are given in units of \unit{\MHz} and are defined as $\delta\nu^\lambda_{i}=\nu^\lambda_i - \nu^\lambda_{48}$. The fermion isotope shifts are relative to the center-of-mass frequency of the hyperfine multiplet. Previous measurements are also given, with references noted in the right column. The measurements obtained in this work have nothing in the reference column.}
    \label{tab:isotope-shift-measurements}
\end{table}

\begin{table}[t]
    \centering
    \fontsize{8.5}{10}\selectfont
    \setlength{\tabcolsep}{1.5pt}

    \begin{tabular}{cccccc}
    
    \hline \hline
    $F$ & $F'$ &
    $\delta\nu_{47}^\mathrm{391nm}$ &
    $\delta\nu_{49}^\mathrm{391nm}$ &
    $\delta\nu_{47}^\mathrm{498nm}$ &
    $\delta\nu_{49}^\mathrm{498nm}$ \\ \hline
    
    $\nicefrac{1}{2}$ & $\nicefrac{1}{2}$ &  & --- &  &  \\
    $\nicefrac{1}{2}$ & $\nicefrac{3}{2}$ &  & --- &  &  \\
    
    $\nicefrac{3}{2}$ & $\nicefrac{1}{2}$ &  & --- &  &  \\
    $\nicefrac{3}{2}$ & $\nicefrac{3}{2}$ & 266.4(4.3)$^{\mathrm{x}}$ & 1401.6(4.2)$^{\mathrm{x}}$ &  &  \\
    $\nicefrac{3}{2}$ & $\nicefrac{5}{2}$ & 26.9(3.0)$^\dagger$ & -861.6(4.2)$^\dagger$ &  & -861.6(4.2)$^{\mathrm{x}}$ \\
    
    $\nicefrac{5}{2}$ & $\nicefrac{3}{2}$ & --- & 1494.6(4.2)$^{\mathrm{x}}$ &  & --- \\
    $\nicefrac{5}{2}$ & $\nicefrac{5}{2}$ & 130.4(3.0)$^\dagger$ & --- &  & --- \\
    $\nicefrac{5}{2}$ & $\nicefrac{7}{2}$ & -208.5(1.8)$^\dagger$ & 924.6(3.1)$^\dagger$ & -1213.7(3.7)$^{\mathrm{x}}$ & -732.9(4.2)$^{\mathrm{x}}$ \\
    
    $\nicefrac{7}{2}$ & $\nicefrac{5}{2}$ & 273.6(2.3)$^{\mathrm{x}}$ & 1392.8(3.5)$^{\mathrm{x}}$ & --- & --- \\
    $\nicefrac{7}{2}$ & $\nicefrac{7}{2}$ & -63.6(1.8)$^{\mathrm{x}}$ & 1059.6(3.0)$^\dagger$ & -958.2(3.0)$^\ddag$ & --- \\
    $\nicefrac{7}{2}$ & $\nicefrac{9}{2}$ & -498.1(2.2)$^\dagger$ & 623.8(2.5)$^\dagger$ & -1023.9(3.1)$^{\mathrm{x}}$ & -537.1(3.3)$^{\mathrm{x}}$ \\
    
    $\nicefrac{9}{2}$ & $\nicefrac{7}{2}$ & 117.9(3.0)$^\dagger$ & 1235.6(2.2)$^{\mathrm{x}}$ & --- & --- \\
    $\nicefrac{9}{2}$ & $\nicefrac{9}{2}$ & -321.0(1.8)$^{\mathrm{x}}$ & 804.1(1.8)$^\dagger$ & -692.3(3.0)$^\ddag$ & -204.9(3.0)$^\ddag$ \\
    $\nicefrac{9}{2}$ & $\nicefrac{11}{2}$ & -849.8(3.0)$^\dagger$ & 271.7(3.0)$^\dagger$ & -770.5(3.0)$^{\mathrm{x}}$ & -279.2(3.1)$^{\mathrm{x}}$ \\
    
    $\nicefrac{11}{2}$ & $\nicefrac{9}{2}$ & -113.9(3.0)$^\dagger$ & 1013.6(2.3)$^{\mathrm{x}}$ & --- & --- \\
    $\nicefrac{11}{2}$ & $\nicefrac{11}{2}$ & -645.2(1.7)$^{\mathrm{x}}$ & 479.6(1.8)$^{\mathrm{x}}$ & -361.1(3.0)$^\ddag$ & 125.8(3.0)$^\ddag$ \\
    $\nicefrac{11}{2}$ & $\nicefrac{13}{2}$ & -1271.5(3.0)$^\dagger$ & -142.4(3.0)$^\dagger$ & -452.8(3.1)$^{\mathrm{x}}$ & 38.4(3.3)$^{\mathrm{x}}$ \\
    
    $\nicefrac{13}{2}$ & $\nicefrac{11}{2}$ & -411.5(3.0)$^\dagger$ & 727.1(3.0)$^\dagger$ & --- & --- \\
    $\nicefrac{13}{2}$ & $\nicefrac{13}{2}$ & -1042.5(1.7)$^{\mathrm{x}}$ & 99.0(2.1)$^{\mathrm{x}}$ & 38.2(3.0)$^\ddag$ & 521.8(3.0)$^\ddag$ \\
    $\nicefrac{13}{2}$ & $\nicefrac{15}{2}$ &  & -623.9(3.0)$^\dagger$ & -67.0(3.0)$^{\mathrm{x}}$ & 419.5(3.0)$^{\mathrm{x}}$ \\
    
    $\nicefrac{15}{2}$ & $\nicefrac{13}{2}$ &  & 372.5(3.0)$^\dagger$ & --- & --- \\
    $\nicefrac{15}{2}$ & $\nicefrac{15}{2}$ &  & -352.6(1.8)$^{\mathrm{x}}$ & 511.8(3.0)$^\ddag$ & 982.5(3.0)$^\ddag$ \\
    $\nicefrac{15}{2}$ & $\nicefrac{17}{2}$ &  &  & 390.4(3.0)$^{\mathrm{x}}$ & 868.1(3.0)$^{\mathrm{x}}$ \\
    
    $\nicefrac{17}{2}$ & $\nicefrac{15}{2}$ &  &  &  & --- \\
    $\nicefrac{17}{2}$ & $\nicefrac{17}{2}$ &  &  &  & 1510.1(3.0)$^\ddag$ \\
    $\nicefrac{17}{2}$ & $\nicefrac{19}{2}$ &  &  &  & 1377.3(3.1)$^{\mathrm{x}}$ \\
    
    \hline \hline
    \end{tabular}
    \caption{Measured hyperfine transitions in the Ti optical pumping (\qty{391}{\nm} wavelength) and laser cooling (\qty{498}{\nm} wavelength) multiplets. $F$ $(F')$ refers to the total angular momentum of the lower (upper) hyperfine level. Line shifts $\nu^\lambda_{i,F\rightarrow F'}$ are expressed as frequencies relative to the $^{48}$Ti transition, with the isotope number and hyperfine levels (transition wavelength) indexed as a subscript (superscript). Shifts are given in \unit{\MHz}. $F$, $F'$ hyperfine transitions that do not occur are left blank, while transition frequencies that were not experimentally measured have a dash. The superscript symbols indicates the method used to measure that line. $^{\mathrm{x}}$ indicates the ``X marks the spot'' method (Fig.~\ref{fig:treasure-map}a,b),  $^\dagger$ indicates the \qty{498}{\nm} wavelength probe was held on resonance while the \qty{391}{\nm} wavelength light frequency was scanned (Fig.~\ref{fig:treasure-map}e), and $^\ddagger$ indicates that three-color spectroscopy was performed to measure the line (Fig.~\ref{fig:treasure-map}d).}
    \label{tab:hyperfine-splittings}
\end{table}

For each line measured by the ``X marks the spot'' method, lock-in amplified 498-nm-wavelength $F\to F+1$ fluorescence spectra were recorded at five distinct 391-nm-wavelength light frequencies, each of which were fit with two Lorentzian features. An overall Gaussian envelope was also included to capture the effect of the transverse velocity distribution of the atomic beam reducing the height of the two peaks as the optical pumping detuning was increased. The positions of the two reonances as functions of the optical pumping detuning were fit to lines, and the crossing point of the fitted lines marked the position of the optical pumping and laser-cooling resonances. This method was also applied to the three bosonic isotopes, $^{46}$Ti, $^{48}$Ti, and $^{50}$Ti, with the measured isotope shifts listed in Table~\ref{tab:isotope-shift-measurements}. The measured $^{48}$Ti resonance was used as the reference frequency for all other measurements.

Once a 498-nm-wavelength $F\to F+1$ hyperfine line was measured in the above procedure, multiple resonances near 319 nm wavelength could be measured by fixing the 498-nm-wavelength probe on resonance and scanning the frequency of the 391-nm-wavelength light. 11 of these spectra were recorded and each was fit with a series of Lorenztian peaks to determine the frequency of the additional lines not measured by the ``X marks the spot'' method.

Finally, three-color depumping spectra were generated by using a double pass acousto-optical modulator to generate the second frequency of light at 498 nm wavelength shown in Fig.~\ref{fig:spectroscopy-schematic}. The RF frequency applied to the AOM was slowly ramped to record the spectrum of the depumping transition. Ten spectra were recorded using this method, and each was fit to a single Lorentzian curve.

\section{King plot analysis}
\label{sec:king-plot-analysis}

 \begin{figure}[h]
    \centering
    \fontsize{8.5}{10}\selectfont
    \setlength{\tabcolsep}{1.5pt}
    \includegraphics[]{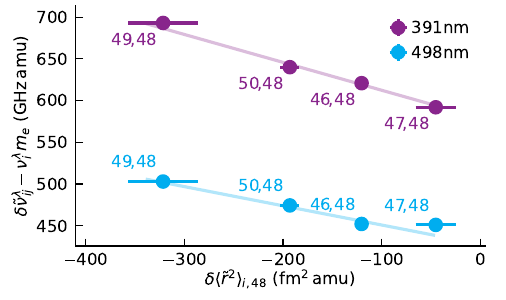}
    \caption{Modified King plots for the optical pumping and laser cooling transitions. Isotope pairs $i$ and $j$ are indicated in text next to each point. The lines show orthogonal distance regression fits of the data to Eq.~\ref{eq:mod-king-eq}.}
    \label{fig:king-plot}
\end{figure}

From the hyperfine multiplet spectra we can also complete a King plot analysis of the isotope shifts on the optical pumping and laser cooling transitions. The isotope shift $\delta \nu^\lambda_{ij}$, observed for a transition $\lambda$, between isotopes $i$ and $j$ is
\begin{align}
    \delta\nu^\lambda_{ij} = \delta\nu^{\lambda,\mathrm{NMS}}_{ij} + k^\lambda\frac{m_i-m_j}{m_i m_j} + F^\lambda\delta\langle r^2\rangle_{ij}
\end{align}
The first term on the right hand side is the so-called normal mass shift (NMS) and is easily calculable as $\delta\nu^{\lambda,\mathrm{NMS}}_{ij}=\nu_i m_e\frac{m_i-m_j}{m_im_j}$, the second term is the specific mass shift (SMS), and final term is the field shift (FS). Within the FS, $F^\lambda$ is an electronic factor associated with the change of the electron charge density at the nucleus, and $\delta \langle r^2\rangle_{ij}$ is the change in mean-square nuclear charge radii between isotope $i$ and $j$. Usually in a King plot analysis, a reference transition for which the SMS and FS are known is used to extract the shifts on the unmeasured transition. For Ti, no such reference transition is available. An alternative King plot analysis can be done if data is available for $\delta \langle r^2\rangle_{ij}$. Furman \emph{et al.}\ have calculated the charge radii changes based on muonic X-ray measurements \cite{furmann_isotope_1996}. Following the standard procedure, we define a modified isotope shift
\begin{equation}
\delta\tilde{\nu}^\lambda_{ij} =
\frac{m_i m_j}{m_i - m_j}\,\delta\nu^\lambda_{ij},
\qquad
\delta\tilde{\langle r^2\rangle}_{ij} =
\frac{m_i m_j}{m_i - m_j}\,\delta\langle r^2\rangle_{ij} .
\end{equation}
The modified King plot relation reads
\begin{equation}
\delta\tilde{\nu}^\lambda_{ij} - \nu^\lambda_im_e =
F^\lambda\delta\tilde{\langle r^2\rangle}_{ij}
+ k^\lambda
\label{eq:mod-king-eq}
\end{equation}
By fitting the modified isotope shifts to a line, $F^\lambda$ can be surmised from the slope and $k^\lambda$ from the intercept. King plots and the resulting fits of the FS and SMS are shown in Fig.\ref{fig:king-plot} and Table \ref{tab:FS-MSMS}.

\begin{table} [h]
    \centering
    \begin{tabular}{cccc}
        \hline \hline
        $\lambda$ (\si{\nm}) & $F^\lambda$ (\si{\MHz/\femto\m^2}) & $k^\lambda$ (\si{\GHz\ amu}) & Ref. \\
        \hline
        391 & \bf{-334(34)}        &  \bf{579(5.2)} &  \\ \hline
        498 & \bf{-230(56)} & \bf{428(9.0)} &  \\ 
         & -186 &  438.8    &  \cite{neely_isotope_2021} \\ \hline \hline
    \end{tabular}
    \caption{Field Shift and Specific Mass Shifts extracted from the King plot analysis shown in Fig.~\ref{fig:king-plot}. Values in bold were determined in this work. The previously reported values for the \qty{498}{\nm} wavelength transition were extrapolated from only two points on the King plot, and therefore reliable uncertainties could not be obtained.}
    \vspace*{\fill}
    \label{tab:FS-MSMS}
\end{table}

\bibliography{references, theory}

\end{document}